\documentclass[preprint,showpacs,preprintnumbers,amsmath,amssymb]{revtex4}

\usepackage{graphicx}% Include figure files
\usepackage{dcolumn}% Align table columns on decimal point
\usepackage{bm}% bold math

\begin{document}
\draft
\title{\bf On the diffraction pattern of C$_{60}$ peapods.}

\author{J. Cambedouzou$^1$, V. Pichot$^2$, S. Rols$^1$, P. Launois$^2$, P. Petit$^3$, R. Klement$^{3,4}$, H. Kataura$^5$ and R. Almairac$^1$}
\address{$^1$Groupe de Dynamique des Phases Condens\'ees (UMR CNRS 
5581),Universit\'e Montpellier II, 34095 Montpellier Cedex 5, France}
\address{$^2$Laboratoire de Physique des Solides (UMR CNRS 8502), Universit\'e Paris
Sud, 91405 Orsay}
\address{$^3$Institut Charles Sadron, 67000 Strasbourg, France}
\address{$^4$Department of Physical Chemistry, Faculty of Chemical and Food Technology, Slovak Technical University, Radlinsk\'eho 9, 812 32 Bratislava, Slovak Republic}
\address{$^5$Nanotechnology Research Institute, National Institute of Advanced Industrial Science and Technology (AIST)
Central 4, Higashi 1-1-1, Tsukuba, Ibaraki 305-8562, Japan}

\date{\today}
%\maketitle
\newcommand{\degree}{$^{\circ}$}%
%\maketitle
\begin{abstract}
%\newline

We present detailed calculations of the diffraction pattern of a powder of
bundles of C$_{60}$ peapods. The influence of all pertinent structural
parameters of the bundles on the diffraction diagram is discussed, which
should lead to a better interpretation of X-ray and neutron diffraction
diagrams. We illustrate our formalism for X-ray scattering experiments performed on peapod samples synthesized from 2 different technics, which present different structural parameters. We propose and test different criteria to solve the difficult problem of the filling rate determination.
\end{abstract}
\maketitle

\vskip 1.5cm

\vskip 1cm

\section{Introduction}

Since their discovery in 1991 \cite{1}, carbon nanotubes have been the
purpose of a large number of studies, dealing both with their mechanical and
electronic properties. In particular, it has been shown that the
intercalation of electron donors or acceptors \cite{2,3,4} into single
wall carbon nanotubes (SWNT) bundles could dramatically modify the
electronic properties of these objects. Rather large molecules are expected
to be inserted into the hollow core of a nanotube that has been shown to present very stable adsorption sites \cite{markj}. C$_{60}$\ is one of the molecules
that have successfully been inserted into SWNT, and a lot of studies have
recently been achieved on the so-called ''peapods''. Those systems
consist of SWNT in which C$_{60}$\ fullerene molecules are inserted 
\cite{6}. Their study stands within the fascinating field of systems in a 
confined geometry \cite{kim,rochefort,9,vavro}. Peapods structural analysis can be performed on a small number of tubes 
(and even on a single tube), 
using transmission electron microscopy (TEM) \cite{6,kataura2001} 
or electron diffraction \cite{hirahara2001,liu}, 
or on macroscopic assemblies,
using Raman spectroscopy \cite{kataura2001,15,kuzmany} 
or X-ray scattering \cite{15,maniwa2003,Zhou}. \newline

Theoretical and experimental studies have already been published on the
diffraction diagram of powder of SWNT bundles \cite{thess,rols}. In particular, the
importance of the distribution of tube diameter on the position of the (10)
Bragg peak in the X-ray and neutron diffraction patterns was pointed
out in ref. \cite{rols,ericprl}. The complete study of the influence of 
all structural parameters of the bundles was performed using a simple
numerical model. It was shown that modeling is essential for a
correct determination of the structural parameters for such inhomogeneous
samples of fairly crystallized nano-objects. Intercalated SWNT bundles form
even more complex nano-crystalline systems. The adsorption sites can be
separated into 3 main locations: inside the tubes, on the outer surface of
the bundles (including the so-called ''grooves''\cite{markj}) and into the
interstitial channels of the 2D triangular lattice. Adsorption of a molecule
into a SWNT bundle can involve modifications of the 2D triangular lattice
(symmetry and/or lattice parameter expansion). These modifications lead to
site-dependent diffraction diagram governed by the structure of the host
(the nanotube bundle), the structure of the adsorbed species inside the bundles and
by crossed interferences between the host and the molecules adsorbed. The
modifications of the diagrams are also found to be radiation dependent \cite
{bendiab}. Therefore, an efficient and correct interpretation of the diffraction
diagram from such complex systems requires the use of simulation.\newline
The study of the structure of C$_{60}$ (and C$_{70}$) peapods by X-ray
diffraction has recently been achieved by Kataura, Maniwa and co-workers
\cite{15,maniwa2003}. The authors could
estimate the C$_{60}$ (and C$_{70}$) filling rates from the
analysis of their measurements. Also, the 1D lattice constant of the C$_{60}
$ chains inside the tubes was found to be smaller than those for 3D
crystals of C$_{60}$. More strikingly, the temperature dependence of the
corresponding feature in the diffraction diagram shows no dependence,
indicating a nearly zero thermal expansion of the C$_{60}$\ chain
inside the tubes, raising questions about possible polymerization of
the C$_{60}$ chains. X-ray diffraction is found to be a very
powerful tool to probe the structure of C$_{60}$\ inside the tubes.
However, further experimental work is needed to understand the properties of
C$_{60}$\ peapods. Such experimental work should include diffraction
investigations on samples showing peapods having various structural
characteristics:

\begin{enumerate}
\item  different tube diameters\

\item  different filling rates\

\item  bundles with various sizes e.g. various numbers of tubes
\end{enumerate}
and particularly, to allow variable degrees of freedom for the fullerene
molecules. Therefore, in this paper, we propose to give a detailed ''step
by step'' and complete study of the diffraction patterns of peapods and we
discuss the characteristic signatures linked to the insertion of C$_{60}$\
inside the SWNTs. We also show how the variation of different structural
parameters, such as tube diameter, filling rate of SWNTs by C$_{60}$, and C$
_{60}$\ adsorption in the outer groove sites at the surface of the bundles
can change the shape of the diffraction pattern. The reader has to keep in mind 
that this report is an
attempt to give experimentalists the necessary tools to characterize their
samples by X-ray and/or neutron diffraction. Very often it appears that unexpected impurity phases are present in SWNT samples, as revealed by a comparison between X-ray and neutron diffraction patterns measured on the same powder. Therefore both techniques are complementary. In the first part of this
paper, we present the theoretical model used to achieve the simulations. We
consider a powder of uncorrelated tubes filled with C$_{60}$\ in
the second part, and the third part deals with powder of bundles of peapods. 
Polymerization of the C$_{60}$ chains is also
considered. We finally compare the
results of our calculations with experimental diffraction patterns. A large part of the discussion is concerned with the determination of the filling rate in the investigated samples.

\section{Principle of the calculations}

In our attempt to reproduce the diffraction pattern of peapods, we developed
a model based on the general equations for x-ray and neutron diffraction in
the kinematical approximation \cite{guinier,16} for which the diffracted
intensity is proportional to the squared modulus of the scattering
amplitude. The latter is defined as the Fourier transform of a configuration
of atoms in the scattering volume. The diffracted intensity thus writes as
follows: 
\begin{equation}
I(\vec{Q})\propto \left| \int_{volume} f_{s}\rho (\vec{r})e^{i\vec{Q}\cdot 
\vec{r}}d^{3}\vec{r}\right| ^{2}  \label{Irho}
\end{equation}

where $\vec{Q}$ is the scattering vector, $\rho (\vec{r})$ is the density of
scatterer at position $\vec{r}$\ in the sample, and $f_{s}$ depends
both on the scattering element (e.g the atomic species) and on the nature of
the incident radiation ($f_{s}$ is a function of the wave-vector 
 modulus Q for X-rays while it is constant for neutrons). Since we
consider the case of a powder experiment, the diffraction is to be averaged
over all the directions of space or, equivalently, over all the orientations
of the scattering vector in the reciprocal space. The mean diffracted
intensity is thus given by: 
\begin{equation}
I(Q)=\frac{\int \int I(\vec{Q})d^2S(\vec{Q})}{4 \pi Q^2}  \label{moypoudre}
\end{equation}
where the integration is performed over the sphere of radius $Q$ and $d^2S(\vec{Q})$ is a surface element of this sphere. Therefore,
the model consists in choosing a convenient -and physically correct-
mathematical form of the density of scatterer $\rho (\vec{r})$. A powerful
approach consists in replacing the discrete carbon atoms by uniformly
charged surfaces for both nanotubes and C$_{60}$ molecules, with a surface
atomic density $\sigma _{c}\sim 0.37 $ atom/\AA$^{2}$.
This value is a little underestimated for C$_{60}$ molecules 
($\sim 0.39 $ atom/\AA$^{2}$) but we take
it as equal for the simplicity. This assumption
results in a loss of information concerning the atomic arrangement at the
surface of the objects forming the sample. It limits the reliability of our
results to Q-values lower than 2 \AA $^{-1}$. 
Below this value, the diffraction pattern
is indeed insensitive to the detailed atomic order. It is however very much
affected in this Q range by the medium range organization e.g. the 2D
hexagonal nanotubes bundles and the 1D C$_{60}$ packing. In the
following, we will therefore be concerned with the two latter levels of
organization in the framework of the homogenous approximation.\newline
The detailed parameters of the model and the definition of the different
variables used thorough this work are presented in Figure 1:

\begin{itemize}
\item  The upper part represents a single tube filled with a linear chain of
C$_{60}$s. Each C$_{60}$-filled nanotube will be considered as a linear
superposition of 1D unit cell, each cell consisting of a cylinder of length
L and of a single C$_{60}$ molecule located at its center.

\item  The lower part represents the peapods organized into bundles e.g. on
a 2D hexagonal lattice, the parameter of which is a function of the
tubes diameter forming it. We introduced a random shift T$_{z}$\ between the
position of the C$_{60}$ molecules on one tube with respect to the corresponding
position on the central tube to avoid unrealistic correlations between the
positions of the C$_{60}$s inside the different nanotubes of the same bundle.
\end{itemize}

\section{Isolated tubes filled with a chain of C$_{60}$ {\bf molecules}: isolated peapods.}

\subsection{Complete filling of the nanotubes}

In this part, we discuss the main features of the diffraction pattern
calculated for a powder of 380 \AA\ long and 13.6 \AA\ large peapods that is
obtained by stacking 40 cylinders of length 9.5 \AA , each cylinder
containing a 3.5 \AA\ radius sphere at its center (see Figure 1). This value of 9.5 \AA\ is not deduced from measurements. It must be considered as a parameter of the model. We
 present in figure 2 the
calculated intensity for this object (bottom), for the tube alone (top) and 
for a C$_{60}$\ chain
alone (middle).\newline
Let I$_{t}$\ be the intensity diffracted by a powder of empty nanotubes. It is a pseudo-periodic oscillating function which is
proportional to the squared modulus of the zero order cylindrical Bessel function J$_{0}$
, as it appears in the following expression of I$_{t}$: 
\begin{equation}
I_{t}(Q)\propto \int_{u=0}^{\pi }\left| A_{t}(\vec{Q})\right| ^{2}sin(u)du
\label{It}
\end{equation}
with A$_{t}$\ the amplitude scattered by the empty nanotube: 
\begin{equation}
A_{t}(\vec{Q})=2\pi Lr_{h}f_{s}\sigma _{c}J_{0}(Qr_{h}sin(u))\frac{sin(Q
\frac{L}{2}cos(u))}{Q\frac{L}{2}cos(u)}\sum_{n=0}^{40}e^{iQnLcos(u)}
\label{Atub}
\end{equation}
and where u, L and r$_{h}$ are defined in figure 1 and where n labels the 1D
unit cells (cylinders of length L).\newline
Let I$_{c}$ be the intensity diffracted by a powder of linear chains of C$_{60}$ molecules: 
\begin{equation}
I_{c}(Q)\propto \int_{u=0}^{\pi }\left| A_{c}(\vec{Q})\right| ^{2}sin(u)du
\label{IC60}
\end{equation}
where A$_{c}$\ is the amplitude scattered by the linear chain of C$_{60}$ molecules: 
\begin{equation}
A_{c}(\vec{Q})=4\pi r_{C_{60}}^{2}f_{s}\sigma _{c}\frac{sin(Qr_{C_{60}})}{
Qr_{C_{60}}}\sum_{n=0}^{40}e^{iQnLcos(u)}  \label{AC60}
\end{equation}
and where $r_{C_{60}}$\ is the radius of a C$_{60}$ molecule (3.5 \AA ). This 
intensity is also a pseudo-periodic oscillating function, but with a
pseudo-period that is nearly twice as long as the one of the nanotubes.
\ Indeed, the radius of a C$_{60}$ \ is nearly half the radial dimension of
the tube. Another interesting feature of the latter curve is the asymmetric
peak at 0.68 \AA $^{-1}$ , which results from the periodic positions of C$
_{60}$s along the linear chain \cite{noteL}.\newline
The first minimum of the intensity diffracted by the whole peapod 
(pointed by an arrow in fig.2) is located
at Q=0.44 \AA $^{-1}$ , whereas it was located at 0.38 \AA $^{-1}$\ in the
empty SWNT diffraction pattern. Following the work on gas adsorption inside
SWNT by Maniwa {\it et al.} \cite{18}, H. Kataura {\it et al.}\ gave a
simple explanation for this feature \cite{15}: the reason 
for the presence of this minimum at this precise value of Q is due to the
sum of the structure factors of the tube and of the C$_{60}$\ which 
equals zero for Q=0.44 \AA $^{-1}$. We will discuss this effect in slightly
different term. Let us write the intensity I$_{p}$(Q)\
diffracted by a peapod as follows: 
\begin{equation}
I_{p}(Q)\propto \int \int \left| A_{t}(\vec{Q})+A_{c}(\vec{Q})\right|
^{2}d^{2}\vec{Q}  \label{amplitudes}
\end{equation}
According to the fact that A$_{t}$\ and A$_{c}$\ may be complex, relation (
\ref{amplitudes}) can be developed as: 
\begin{equation}
I_{p}(Q)\propto I_{t}(Q)+I_{c}(Q)+2\int \int \left[ Re(A_{t}(\vec{Q}
))Re(A_{c}(\vec{Q}))+Im(A_{t}(\vec{Q}))Im(A_{c}(\vec{Q}))\right] d^{2}
\vec{Q}  \label{developpement}
\end{equation}

where Re(A$_{t}$) and Im($A_{t}$) stand for the real and imaginary parts of A$_{t}$, respectively. Relation (\ref{developpement}) can be rewritten as: 
\begin{equation}
I_{p}\propto I_{t}+I_{c}+I_{CI}  \label{xterm}
\end{equation}
If attention is given to figure 2c, it is clear that the profile of I$_{p}$
is quite different from (I$_{t}$+I$_{c}$). This latter quantity is indeed
the signature of a sample containing a tube decorrelated from a linear chain
of C$_{60}$. In a peapod sample, there is a strong correlation in the
relative positions of the tube and the chain of C$_{60}$, since the C$_{60}$s\ 
are located inside the tube. This correlation is revealed by
the presence of the crossed interference term I$_{CI}$\ which can be
positive or negative. At Q=0.44 \AA $^{-1}$, the crossed term compensates
 the (I$_{t}$+I$_{c}$) term, lowering I$_{p}$\ to zero or nearly zero \cite{notezero}. This value for Q is accidently the same as that of the (10) peak position for
a bundle made of empty SWNTs stacked into a 2D hexagonal lattice. This
coincidence will cause dramatic changes in the diffraction pattern
relative to peapods bundles (see part IV).\newline
The increase of the tube diameter induces a decrease of the oscillation
period as it is expected when considering larger diffracting object (see figure 3). The C$
_{60}$\ periodicity characteristic peak at 0.68 \AA $^{-1}$, as well as its
first harmonic at 1.36 \AA $^{-1}$, are observable in all the calculated
diffraction profiles. For peapods of radius r=5.42 \AA  \space(for example C$_{60}$@(8,8) peapods) the first peak is
not separated from the first oscillation, but is clearly visible. The same
behavior is observed for r=8.1 \AA\ peapods (for example C$_{60}$@(12,12) peapods) on the second oscillation,
but the peak amplitude is weaker. In fact, the larger the tube diameter and
the weaker this peak. The latter effect can be easily understood: an
increase of the tube diameter implies an increase of the tube surface which
leads to a preponderance of the response of the tubes over the
response of the C$_{60}$\ chains, in which the number of scatterers remains unchanged. The cases of C$_{60}$@(8,8) and C$_{60}$@(12,12) peapods are discussed here as extreme cases for the influence of the tube diameter. However, it must be pointed out that the insertion of C$_{60}$ in a (8,8) peapod is unlikely due to the too small diameter of the tube, and C$_{60}$s inside a (12,12) peapod would possibly lead to zigzag or helical chains in place of linear chains, since the center of the tube is no longer the most favorable location for C$_{60}$ regarding to van der Waals interactions.\newline
The bottom part in figure 3 reveals the consequences of a change in the
inter-C$_{60}$\ length L inside the linear chain. A downshift of the
characteristic peak of C$_{60}$s periodicity is evidently observed when L
increases, but we also note a reinforcement of its intensity. One remarks a
striking effect: although the relative density of C$_{60}$\ increases, the
characteristic peak relative to the C$_{60}$\ periodicity strongly
decreases. This is due to the multiplication factor arising from the 
intensity diffracted by a C$_{60}$ molecule, which
reaches its first zero at 0.9 \AA $^{-1}$, as shown in figure 2. The closer
the peak position to this value, the weaker the peak. One must be careful
with the fact that the respective proportion between tube and C$_{60}$s is
no longer the pertinent parameter to account for the explanation of the
changes in the peak intensities. Thus, it may prove dangerous to focus only
on the relative intensity of this peak, and it seems necessary to consider
the whole diffraction pattern to derive reliable structural information.
\newline
 In several cases, the use of an analytical formula can prove more comfortable than 
the method presented above. For infinite tubes, numerical
calculations are performed after the limit of infinite tube length\ was
taken in the above equations, leading to the following equation
(demonstrated in appendix 1 eq. \ref{app1isole}), where the first term comes from the Fourier
transform of the structure projected in a plane perpendicular to the tube
axis, while the second one comes from the periodicity of the C$_{60}$
chains:\  
\begin{equation}
I_{p}(Q)\propto \frac{f_{s}^{2}}{Q}\left( (2\pi Lr_{h}\sigma
_{c}J_{0}(Qr_{h})+4\pi r_{C_{60}}^{2}\sigma _{c}\frac{sin(Qr_{C_{60}})}{
Qr_{C_{60}}})^{2}+2Int(\frac{QL}{2\pi})(4\pi r_{C_{60}}^{2}\sigma _{c}\frac{sin(Qr_{C_{60}})}{
Qr_{C_{60}}})^{2}\right)   \label{powderNTsinfinite}
\end{equation}
 In this expression $Int(\frac{QL}{2\pi})$ is the integer part of\ ($\frac{ QL}{ 2\pi }$ ).
\newline

\subsection{Partial filling of the nanotubes}

In the case of a real sample, it seems reasonable to consider that all tubes
may not be fully filled with C$_{60}$s, despite efforts made to obtain
filling rates as high as possible \cite{kataura2001}. Two different
hypotheses are discussed here.\newline
The first hypothesis is to consider a random filling 
of the tubes without assuming clustering effects 
inside the tubes : the molecules are randomly positioned within a tube, the only constraint 
being a minimum distance of L between them (one may note that it implies that in the limiting case
of 100\% filling, the molecules necessarily form ordered chains). Calculations are performed 
within the finite tube length model, for a given filling ratio of the nanotubes. 
The effects of a random incomplete filling of the nanotubes are presented in the upper part 
of figure 4. One observes the vanishing of the C$_{60}$-C$_{60}$ characteristic peak  with decreasing filling
rates; it completely disappears for filling rates below 85\%. If we compare the shapes of the 85\% 
and the 100\% diffraction profiles, one finds that the minimum at 0.44 \AA$^{-1}$ 
for the 85\% filled sample no longer lowers to zero. 
The additional intensity can be attributed to the effect of the 
disorder induced by the random filling of the nanotubes by the C$_{60}$ molecules. Moreover, the 
first minimum in
intensity shifts to lower Q values for decreasing rate of fullerenes, which
can be explained as in figure 2 by compensation effects between tube,
fullerene and interference terms. \newline
As a second hypothesis, one can consider the partial filling of the tubes with long
('quasi-infinite') chains of C$_{60}$. We assume here that the
molecules tend to cluster within nanotubes.\ Indeed, this should correspond
to a low energy configuration of the system, the energy being lowered by the
attractive C$_{60}$-C$_{60}$\ interactions.\ \ This hypothesis is
supported by observations reported in ref.\ \cite{hirahara2001}. Calculated
diffraction patterns for different filling rates (50\%, 75\% and 100\%) are
presented in the lower part of figure 4. Here we use the infinite tube length model 
(detailed calculations are given in
appendix 2). As it was already mentioned in the first hypothesis, we observe the vanishing of
the C$_{60}$-C$_{60}$ characteristic peak at low filling rates, but it interestingly  
disappears here for
a filling rate of 50\% which is much lower than the 85\% observed in the first part.  
This is due to the long chains assumption : the C$_{60}$-C$_{60}$ distance is preserved
for the different filling rates. Another feature of these diffraction patterns 
is that the first minimum goes to zero for all filling rates, contrarily to what was found 
within the first hypothesis, which is due to lower disorder. \newline
If one takes into account the differences pointed out between the 2 hypothesis 
discussed here, 
one should in principle be able to determine the way a sample of isolated peapods 
is filled.
Unfortunately, the Q range around 0.44 \AA$^{-1}$ is usually perturbed by parasitic 
signals (intense scattering at small wave-vectors ), so it is 
very difficult to use the value of the first minimum of intensity as a clue to 
determine the filling mode. As a result, the C$_{60}$-C$_{60}$ characteristic peaks 
remain the only observable features for the estimation of the filling rate and of the filling mode. 
An important result from our calculations is that C$_{60}$ molecules in isolated peapods 
with a random filling rate below 85\% and 
peapods with a long chain 
filling rate under 50\% are undetectable by diffraction.

\subsection{Polymerized C$_{60\text{ }}$molecules inside nanotubes}

One of the remarkable properties of the C$_{60}$\ molecule is its ability to
form covalent bonds through quite different routes. In crystalline C$_{60}$, 
polymers can be obtained (i) by photopolymerization, (ii) under
pressure and at high temperature , (iii) in doped samples \cite
{sundqvist,forro}. Photo-induced \cite{15,kataura2001} and charge
transfert induced \cite{pichler2001} polymerization have also been demonstrated
in C$_{60}$\ peapods, using Raman spectroscopy.\ It is interesting to
consider the effect of polymerization on diffraction patterns.\ Let us
consider peapods where the chains can be formed of n-polymers (n=1:
monomers, n=2: dimers, n=3: trimers...).\ For the sake of simplicity, we
deal with the case of completely filled nanotubes. The distance L$_{b}$\
between bonded C$_{60}$ molecules is taken to be 9.2\AA\ as in crystalline
polymerized samples, and the distance L between unbonded molecules is fixed
at 9.5\AA\ as above.\ The distance between monomers is smaller in
peapods than in crystalline C$_{60}$, where it is equal to about 10\AA ,
possibly because of interactions with the nanotubes. However, there is no
reason to take a smaller value for the distance between bonded molecules
because it is mainly determined by the covalent bonding between them.\ It is
shown in appendix 3 that the scattered intensity of a powder of peapods
filled with chains of n-polymers writes:

\begin{displaymath}
I_{p}(Q)\propto \frac{f_{s}^{2}}{Q}( (2\pi r_{h}(L+(n-1)L_{b})\sigma_{c}
J_{0}(Qr_{h})+4\pi r_{C_{60}}^{2}\sigma _{c}n\frac{sin(Qr_{C_{60}})}{
Qr_{C_{60}}})^{2} %\nonumber\\
\end{displaymath}
\begin{equation}
+2(1-\delta _{M,0})\sum_{k=1}^{M}(4\pi r_{C_{60}}^{2}\sigma
_{c}\frac{sin(Qr_{C_{60}})}{Qr_{C_{60}}}
\frac{\sin (k\pi
nL_{b}/(L+(n-1)L_{b}))}{\sin (k\pi L_{b}/(L+(n-1)L_{b}))})^{2})
%\right
\label{polym}
\end{equation}
\ where the second term appears only for Q values such that M -the integer
part of ($\frac{Q(L+(n-1)L_{b})}{2\pi}$) - is not zero. Calculated patterns
for n=1, 2 and 3 are drawn in fig.5. At each ${ Q=k2}\pi /(L+(n-1)L_{b})$\ value (with k integer),
asymmetric peaks characteristic of the chain periodicity can be observed.
Despite the change of the period -equal to (L+(n-1)L$_{b}$) - with n, the
spectra look quite similar.\ It can easily be explained from eq. \ref{polym}
: the asymmetric peak intensity is multiplied by $\left[ \frac{\sin (k\pi
nL_{b}/(L+(n-1)L_{b}))}{\sin (k\pi L_{b}/(L+(n-1)L_{b}))}\right] ^{2}$\ ,
which is close to zero except for k=n, 2n,... The first intense peak of the
n-polymer diffraction pattern is thus located at Q=n${ 2}\pi
/(L+(n-1)L_{b})$\ : Q=0.672\AA $^{-1}$\ for dimers, 0.676 \AA $^{-1}$ for
trimers, to be compared with 0.661\AA $^{-1}$\ for monomers.\ The upper
value, corresponding to infinite polymers, is Q=2$\pi /9.2=$0.683\AA $
^{-1}.\;$

In summary, scattering analysis of polymerization of C$_{60\text{ }}$
molecules in peapods samples should be based on a careful study of the
position Q$_{0}$\ of the first intense asymmetric peak, and on the search
for lower intensity peaks at kQ$_{0}$/n (n=2,3...; k=1 to (n-1)) to identify
n-polymers.

{\bf \ }

\section{Bundles of peapods.}

\subsection{Complete filling}

Peapods are packed into bundles where they are maintained together by van
der Waals inter-tubes interactions. This organization is clearly visible on TEM pictures \cite{15,kataura2001,maniwa2003}. In this part,
we calculate and discuss the diffraction pattern for such objects. Calculations are performed both for 
peapods of finite length  and for peapods of infinite length (as detailed in appendices 
1 and 2).  The relative positions of the C$_{60} $ chains
 along the tube axes (T$_{z}$ in fig.1) are assumed to present no correlation 
from one tube to  another. Indeed, a C$_{60} $ chain
interacts the most with the nanotube in which it is located, and nanotubes within a bundle present 
different helicities \cite{henrard2000}.
 \newline
Figure 6 shows the comparison between the diffraction profiles
calculated for a powder of bundles of 12 empty nanotubes and that of bundles of 12
peapods, with different tube radii. In all cases, we consider 380 \AA\ long nanotubes organized on a 2D hexagonal lattice with a 3.2 \AA\ van der Waals length between 2 adjacent nanotubes. Let us first consider the upper part of figure 6, which deals with (10,10) tubes and peapods, of radius r=6.8 \AA. For peapods, the additional peaks characteristic of the 1D periodicity of the C$_{60}$ chains are 
indicated by an arrow.
We remark that most of the characteristic peaks observed in
the diffraction pattern of the empty SWNT bundles show up in the diffraction
pattern of the peapod bundles, except those located in the low Q range where
a lack of intensity is obtained for the peapods. 
An important difference between the pattern of the peapod bundles and the empty
nanotube bundles is thus the disappearance of the (10) Bragg peak at 0.44 \AA $
^{-1}$. This peak (of finite width because of the small bundle size) is replaced 
by a minimum delimited by two smaller peaks. This can be explained on the basis of 
 part III results, since we observed that the intensity diffracted by a single peapod
lowers to zero  at 0.44 \AA $^{-1}$, which is around the position of the
(10) lattice peak of the bundles. This important extinction process in the case of SWNT 
bundles has already been found
and discussed for different kinds of intercalated molecules like gas
molecules \cite{15,fujiwara2001} or in iodine doped nanotubes samples \cite
{bendiab}. Middle and lower parts of figure 6 show that the extinction phenomenon is modified when the tube diameter is changed. For example, we consider the case of bundles of 12 empty nanotubes and 12 peapods of radius r=5.42 \AA  \space((8,8) tubes, middle part of figure 6) and r=8.1 \AA  \space((12,12) tubes, lower part of figure 6). One can see that the (10) peak is not split into two parts for the bundle of 
(8,8) peapods, but appears shifted to lower Q values. By contrast, this peak completely disappears in the case of the bundle of (12,12) peapods. When tube is thinner or larger in diameter, the progressive
loss of the accidental adequate conditions implies the extinction phenomenon
to be lost. \newline
It is important to be aware that the extinction can occur for values of Q
that are slightly different from the (10) peak position. In that case the
(10) peak does not appear split, but seems shifted because only one side of
the peak is lowered. Therefore, a direct interpretation of such apparent
shift in terms of a change of the lattice parameter is inappropriate, and
direct conclusions about structural changes based on the observation of the
(10) peak alone, prove very hazardous. The analysis of peapods diffraction patterns is consequently not straightforward and should 
be based on comparison between measurements and calculations.

\subsection{Partial filling}

As for isolated tubes, all tubes may not be fully filled with C$_{60}$s. 
In this part, 4 different filling modes are discussed.\newline
The first case (case a) consists in a random filling, where each tube of the bundle is
filled as described in part III.B.
Calculations are performed 
within the finite tube length model.
In the 3 other cases, which are treated within the infinite tube assumption 
(detailed calculations are given in appendix 2), the C$_{60}$s are all stacked into long chains, 
but the way these chains are distributed into the bundles changes with the case. One can indeed 
consider an homogeneous filling (case b), where the tubes are all filled with the same number of 
C$_{60}$ molecules, or an inhomogeneous filling (case c), where the filling rate of each tube 
of the bundle is slightly different. The last case to be discussed (case d) consists in a 
mix of fully filled bundles and empty bundles (see the right part of figure 7).\newline
We present the results for filling rates of 85\% and 50\% for bundles of 12 nanotubes of radius r=6.8\AA\ in the left part of figure 7. The main difference between all diffraction patterns calculated for a 85\% filling rate and all those calculated for a 50\% filling rate is visible in the low Q range. The (10) Bragg peak is indeed still splitted and almost invisible at the 85\% filling rate, while it clearly reappears at the 50\% filling rate. This observation can be explained by the fact that the accidental conditions allowing the extinction of the (10) Bragg peak are progressively lost when the proportion of C$_{60}$ decreases in the sample.\newline
If more attention is given to the a), b), c) and d) diffraction profiles for a given filling rate, other features linked to the different filling modes can be extracted from the figure. One can first consider the Q range below 0.6 \AA$^{-1}$. The intensity in this region is the least for the b) configuration, which is the least disordered configuration. If we compare with the inhomogeneous filling (mode c), and then with the random filling (mode a), we observe a progressive increase of the intensity in the low Q range, corresponding to the progressive increase of the disorder in the system, as already mentionned in part III.B. The d) case must be considered separately from the other cases because the intensity in the low Q range is here the sum of the intensities of 2 decorrelated systems ( full peapods and empty nanotubes). 
\newline
One can also focus on the C$_{60}$-C$_{60}$ characteristic peak. 
This asymmetric peak is clearly visible on the diffraction profiles of the b), c) and d) filling modes,
 even for the 50\% filling rate, whereas it never appears at this rate for isolated peapods. 
In addition, this peak is invisible at both filling rates in the a) filling mode. 
Thus, the observation of such a feature in an experimental diffraction pattern should stand as the signature of a long chain organization of the C$_{60}$ molecules inside the tubes.

\subsection{Chains of C$_{60}$ molecules in the outer groove sites of the bundles}

We finally discuss the effects induced by the filling of the channels
located at the bundle surface with linear chains of C$_{60}$s , for bundles
of various size (figure 8). All tubes have a radius of 6.8 \AA. When we consider the diffraction pattern of a 6
peapods bundle where the external channels have been completely filled, we
note 2 main differences with regards to a naked peapod bundle.
First, the peak at 0.68 \AA $^{-1}$\ is significantly reinforced due to the
additional C$_{60}$ molecules in the external channels. Secondly,
the intensity in the low Q range as well as between 1 and 1.5 \AA $^{-1}$
appears strengthened. As it is shown in figure 2, the C$_{60}$\ response is
high in these Q ranges, so this effect is just due to the increase of the C$
_{60}$\ proportion in the sample. It is evidently clear that the higher the
number of tubes in the bundle, the weaker the effect described just above.
The reason for that lies in the simple fact that volume grows faster than
surface when the size of the bundle increases. According to these results,
we are able to conclude that an experimental diffraction pattern with a high
intensity in the 1 to 1.5 \AA $^{-1}$\ Q range may be relative to a sample made
of small sized bundles, saturated with C$_{60}$s in both the inner space of
the tubes and in the external channels.\newline

\subsection{Determination of the filling rate}

In this section we try to define a method to determine the filling rate from experimental data. 
Two criteria are proposed which are based on the analysis of the results obtained from the model. 
As it was explained previously, the (10) Q range is strongly perturbed, so this Q range is disregarded.
 The first criterion consists to remark that the peak centered around 0.7 \AA $^{-1}$\ contains 
two contributions (fig. 7):     i) the response of the C$_{60}$ lattice at 0.663 \AA $^{-1}$\ and 
ii) the (20) reflection of the bundle lattice centered at 0.706 \AA $^{-1}$\ . 
The ratio of the two respective amplitudes is expected to vary as a function of the filling rate.  
However two problems arise. The first one concerns the type of filling which is unknown. 
The second concerns the background determination. We assume that the background 
can be represented by a straight line between 0.6 \AA $^{-1}$\ and 0.8 \AA $^{-1}$\ 
as is shown in the inset of fig. 9a. In fig. 9a the amplitude ratio is drawn 
as a function of the filling rate and for the two extrem types of filling, i.e. random 
and homogeneous (a) and b) cases, respectively). Cases c) and d) have intermediate behaviors, so they are not presented in this figure.
  Thus for a given experimental amplitude ratio (on vertical axis) one gets a range 
for the filling rate.\newline
The second criterion consists to use: i) the amplitude of the C$_{60}$ 1-D lattice peak around 
1.3 \AA $^{-1}$\ as this peak arises in a Q range where the C$_{60}$ contribution is strong, 
and ii) the amplitude of the (21) reflection of the bundles around 1.1 \AA$^{-1}$\ as the C$_{60}$ 
contribution is small in this Q range. The calculated diagrams of fig. 7 
were analyzed with a 3 peaks model from which we extracted the two amplitudes (fig. 9b). 
Again the background estimation has a large incidence on the results. \newline 
The second criterion seems to be more reliable as it presents a smaller dependence 
on the type of filling. It is of easier use. However it has been established for SWNT 
bundles corresponding to diameters occuring in the case of  electric arc or laser synthesis, 
i.e. centered around 1.4 nm and with a small dispersion in diameter 
distribution (FWHM $\leq  2$ \AA\ ). 
In the next section these criteria are applied 
to the estimation of the filling rate in two different samples.
\newline

\section{\bf Comparison between calculated and experimental diffraction patterns}

Diffraction patterns were performed using a powder diffractometer equiped with
a curved position sensitive detector INEL CPS120 allowing one to measure simultaneously a
range of 2$\theta $ angles from 2 to 120$^{\circ }$. A wavelength of 1.542
\AA\ was used. We measured the
diffraction pattern of two different peapod samples (powder sample or numerous small pieces of bucky paper in 
glass capillaries). 
Sample A (figure 10a) was synthesized as follows : the SWNT material was first soaked
 in concentrated nitric acid (HNO$_{3}$ 65\%) and sonicated for approximatively 30 min, then treated in boiling HNO$_{3}$ at 140-150 $^{\circ }$C for 4 hours. After sedimentation of the acid treated SWNT material in distilled water, SWNTs were separated from the acidic supernatant by centrifugation. The acid treated material was washed several times with distilled water (several centrifuge-washing-decantation cycles) then washed twice with ethanol. Finally, the SWNT material was dried under low pressure during overnight. The dried acid treated and washed SWNTs were subsequently heated in air at 420-430 $^{\circ }$C for 30-40 min. After oxidation, the material was mixed with an excess of C$_{60}$ powder in a glass tube. The glass tube was sealed under high vacuum, then heated at temperatures 550-600 $^{\circ }$C for 72 hours. After heating, the ampoule was cooled down to room temperature and opened. The excess of C$_{60}$ was removed by washing the SWNT material with toluene. Finally, the material was washed once with ethanol and dried under reduced pressure for overnight. The preparation method of sample B (figure 10b) has been described previously \cite{kataura2001}.
\newline  
The diffraction patterns of these two samples look quite different. We note the presence 
of a broad underlying structure from 1 to 2.2 \AA $^{-1}$ 
in the diffraction patterns of sample A (dotted line in
fig. 10a). 
Applying the two criteria defined in part IV.D to the determination of the filling rate (after substracting 
the dotted line 
for sample A), one obtains
 75 to 95\% (using the first method) compared to 85\% (second method) for sample A, 
and 77 to 96\% (first) compared to 95\% (second) for sample B (see the horizontal lines in fig. 9). 
However, as is shown below, if the criteria can be applied to sample B where characteristic C$_{60}$ peaks are 
clearly observed, 
they only give an estimation of a upper value of the filling rate for sample A. 
\newline
Diffraction profiles were fitted by optimized calculations.
Calculated diffraction patterns were convoluted with a convenient resolution 
function in order to be compared with the experimental profiles.
We also introduced a distribution of tube diameter in the calculation. The
latter was considered to be of Gaussian shape \cite{rols,ericprl}.\newline
Concerning sample A,  the comparison with the pristine SWNT
powder allows us to put forward the following main characteristics of the
peapod diffraction profile: a weak (10) peak, an enlargement of the second
peak at its low Q-side (0.68 \AA $^{-1}$) and additional intensity around
1.3 \AA $^{-1}$.  We obtained the best fit of the diffraction pattern considering a structure
made of small bundles of 6 peapods, with a large
tube radius distribution centered around 6.8 \AA\ and chains of C$_{60}$s in the external channels of the bundles. 
Considering the 4 filling modes described above, a very high filling rate implies a sharp peak at 0.68 \AA$^{-1}$, 
standing for the C$_{60}$-C$_{60}$ periodicity. Such a sharp peak is not observed in the diffraction pattern, 
so we introduced a distribution of C$_{60}$-C$_{60}$ lengths from 9 to 10 \AA\  into the model, 
leading to an improvement of the fit. 
This distribution could testify to the presence of mixed dimer, trimer or n-polymers chains of C$_{60}$ 
into the tubes in this sample.
However, if by-products of the
chemical treatment were present inside nanotubes with the C$_{60}$ molecules, this would
explain the decrease of the (1,0) peak intensity together with 
an enlargement of the C$_{60}$ periodicity peak since periodicity would be perturbated. We should thus note here that
on the basis of the present results, the filling rate values determined here, above 75\%,  might be over-estimated.
\newline
Concerning sample B, the general shape of the diffraction profile is better
fitted. Different kinds of filling modes have been studied in order 
to give a reasonnable range for the filling rate. If a d) filling mode is chosen, 
we obtain the plain line of figure 10b, and a filling rate of about 75\%.  
The model includes 30 tubes per bundle with a very narrow
distribution of tube radii centered at 6.76 \AA\ (same radius as in ref. \cite{maniwa2003}), 
and an inter-C$_{60}$ length of 9.8 \AA.  The good agreement between the
calculation and the rest of the diffraction pattern is the proof of a
reliable characterization of the sample and of a high quality process of
synthesis.\newline
The diffraction patterns measured for both samples show up a clear difference in their respective structures. 
Clear features in the diffraction pattern of sample B allow an unambiguous characterization and a 
reliable estimation of its filling rate. Furthermore, our results are in good agreement with what was previously determined for this sample in other x-ray diffraction and transmission electonic microscopy studies.\cite{15} On the other hand, one must be more careful with the determination 
of the filling rate of sample A,  as noted above.

\section{Conclusion}

We presented in detail the formalism to calculate the
diffraction diagram of peapods at different level of organization: isolated
and organized into bundles. This formalism allows to numerically
investigate the main characteristics of the diffraction patterns of peapod
samples, and to discuss those driving to a pertinent characterization of
real samples. In particular, the isolated peapod study shows how the
concentration of C$_{60}$s inside the
tubes can shift the positions of the diffracted intensity zeros, and
consequently lead to an accidental extinction of the (10) Bragg peak in the
diffraction profile of peapod bundles, whereas there is no change in the
bundle arrangement. These results show that one must be extremely careful about the
correct interpretation of the changes in position and intensity of the (10)
Bragg peak in experimental data concerning peapods and also all the inserted
samples of SWNTs. Furthermore, we saw that the diffraction pattern of peapods has to be considered in its entire 0 to 2 \AA$^{-1}$ Q range in order to derive reliable characterization of the sample. In particular, much attention has to be paid  to both the intensity and shape of the C$_{60}$-C$_{60}$ characteristic peaks. 

Those features are discussed in the measured diffraction
patterns of two different samples.

\section{Acknowledgements}

It is our pleasure to acknowledge J.-L. Sauvajol and P.A.\ Albouy for
fruitful discussions. H. Kataura acknowledges for a support by Industrial Technology Research Grant Program in '03 from New Energy and Industrial Technology Development Organization (NEDO) of Japan.

\section{Appendix 1}

The calculation of the scattering pattern from a powder of peapods is
detailed in this appendix for the case of nanotubes of infinite length.

For a given wavevector $\vec{Q}$  the intensity coming from an isolated peapod is:

\[
I(\vec{Q})=
{\displaystyle {F(\vec{Q})F^{*}(\vec{Q})}}\]

The form factor of a peapod of length L, 
containing one C$_{60}$
molecule at half height, writes

\begin{equation} 
F(\vec{Q})=f_{s}(2\pi r_{h}L\sigma _{c}J_{0}(Q_{//}r_{h})
{\displaystyle {\sin (Q_{z}L/2) \over Q_{z}L/2}}
+4\pi r_{C_{60}}^{2}\sigma _{c}
{\displaystyle {sin(Qr_{C_{60}}) \over Qr_{C_{60}}}}
) \label{formfactor}
\end{equation} 
In this expression $\overrightarrow{Q_{//}}$\ is the projection of the wave-vector $\vec{
Q}$\ perpendicularly to the nanotube axis and Q$_{z}$\ is its projection
along the axis.

For a peapod of length $N_cL$, it becomes

\begin{equation} 
F(\vec{Q})=f_{s}(2\pi r_{h}L\sigma _{c}J_{0}(Q_{//}r_{h})
{\displaystyle {\sin (Q_{z}L/2) \over Q_{z}L/2}}
+4\pi r_{C_{60}}^{2}\sigma _{c}
{\displaystyle {sin(Qr_{C_{60}}) \over Qr_{C_{60}}}} )
\sum 
_{n=0}^{N_c-1}\exp (iQ_{z}nL)
 \label{formfactorNL}
\end{equation}

The intensity per length unit thus writes 
\begin{displaymath} 
I(\vec{Q})= f_{s}^{2}[2\pi r_{h}L\sigma _{c}J_{0}(Q_{//}r_{h})
\frac {\sin (Q_{z}L/2)}{Q_{z}L/2}
\end{displaymath} 
\begin{equation} 
+4\pi r_{C_{60}}^{2}\sigma _{c}
\frac {sin(Qr_{C_{60}})}{Qr_{C_{60}}}
]^{2}
\frac {1}{N_cL}
\sum 
_{n=0}^{N_c-1}\exp (iQ_{z}nL)
\sum 
_{m=0}^{N_c-1}\exp (-iQ_{z}mL) \label{ideQ}
\end{equation}

Now we consider an infinite tube: $N_c \rightarrow \infty $.
 Using the relation 
\begin{displaymath}
\lim_{N_c\rightarrow \infty }
{\displaystyle {1 \over N_cL}}
\mathop{\displaystyle \sum }
_{n=0}^{N_c-1}\exp (iQ_{z}nL)
\mathop{\displaystyle \sum }
_{m=0}^{N_c-1}\exp (-iQ_{z}mL)=\mathop{\displaystyle {2\pi  \over L^2}}
\mathop{\displaystyle \sum }
_{k=-\infty }^{\infty }\delta (Q_{z}-2\pi k/L)
\end{displaymath}
where $\delta{(x)}$ is the Dirac distribution and where $k$ is an integer, it follows 
that the intensity per unit length scattered by an isolated peapod of infinite
length writes

\[
I(\vec{Q})=f_{s}^{2}[2\pi r_{h}L\sigma _{c}J_{0}(Q_{//}r_{h})
{\displaystyle {\sin (Q_{z}L/2) \over Q_{z}L/2}}
+4\pi r_{C_{60}}^{2}\sigma _{c}
{\displaystyle {\sin(Qr_{C_{60}}) \over Qr_{C_{60}}}}
]^{2}   {\displaystyle {2\pi \over L^2}}
\mathop{\displaystyle \sum }
_{k=-\infty }^{\infty }\delta (Q_{z}-2\pi k/L) 
\]
which can be written as:

\begin{displaymath}
I(\vec{Q})=f_{s}^{2} {\displaystyle {2\pi \over L^2}}[(2\pi r_{h}L\sigma _{c}J_{0}(Q_{//}r_{h})
{\displaystyle {\sin (Q_{z}L/2) \over Q_{z}L/2}}
+4\pi r_{C_{60}}^{2}\sigma _{c}
{\displaystyle {sin(Qr_{C_{60}}) \over Qr_{C_{60}}}}
)^{2}
\delta (Q_{z})
\end{displaymath}
\begin{equation}
+(4\pi r_{C_{60}}^{2}\sigma _{c}
{\displaystyle {sin(Qr_{C_{60}}) \over Qr_{C_{60}}}}
)^{2}
\mathop{\displaystyle \sum }
_{k\neq 0}\delta (Q_{z}-2\pi k/L)] 
 \label {I1tube}
\end{equation}

The $\delta (Q_{z})$ dependent term \ is the Fourier transform of the
structure projected on a plane perpendicular to the nanotube axis, while the 
$\delta (Q_{z}-2\pi k/L)$ dependent term comes from the periodicity of the C$
_{60}$\ chain.

Now we use eq. \ref {moypoudre} to calculate powder average. If $g(Q_{//},Q_z)$ represents the factor multiplicating $\delta (Q_{z})
$ in the previous expression, the integration of the $\delta (Q_{z})$ dependent term over the angles $u$ and $\varphi $ gives
\[
\int_{0}^{2\pi} d\varphi \int_{u=0}^{\pi } g(Q_{//},Q_z) \delta(Q\cos(u)) \sin(u)  du = {\displaystyle {2\pi} \over Q} g(Q,0)
\]

where $Q_z=Qcos(u)$. The integration of the $\delta (Q_{z}-2\pi k/L)$ dependent term reduces to
\[
\mathop{\displaystyle \sum }
_{k\neq 0}\int_{0}^{2\pi} d\varphi \int_{u=0}^{\pi } \delta (Q\cos(u)-2\pi k/L) \sin(u)  du 
\]
which is equal to $[2{\displaystyle {2\pi} \over Q} Int(QL/2\pi)]$ . Here $Int(QL/2\pi)$ 
is the integer part of ($QL/2\pi$) : the asymmetric shape of the peaks characteristic 
of the C$_{60}$ 1D periodicity can be understood through this formula (the sawtooth line shape).

It follows that the intensity scattered by a powder of \textit{isolated} peapods is given by

\begin{displaymath}
I_{p}(Q) = \frac{(2\pi)^2 f_{s}^{2}}{QL^2}[ (2\pi r_{h}L\sigma
_{c}J_{0}(Qr_{h})+4\pi r_{C_{60}}^{2}\sigma _{c}\frac{sin(Qr_{C_{60}})}{
Qr_{C_{60}}})^{2}
\end{displaymath}
\begin{equation}
+2Int(QL/2\pi)(4\pi r_{C_{60}}^{2}\sigma _{c}\frac{sin(Qr_{C_{60}})}{
Qr_{C_{60}}})^{2}]  \label{app1isole}
\end{equation}

Let us now calculate the intensity scattered by a powder of peapod \textit{bundles}. 
The expression of the form factor of a bundle of peapods of length $N_cL$ is :

\begin{displaymath}
F(\vec{Q})=f_{s}\sum_{i} [ 2\pi r_{h}L\sigma _{c}J_{0}(Q_{//}r_{h})
{\displaystyle {\sin (Q_{z}L/2) \over Q_{z}L/2}}
\end{displaymath}
\begin{equation}
+4\pi r_{C_{60}}^{2}\sigma _{c}
{\displaystyle {sin(Qr_{C_{60}}) \over Qr_{C_{60}}} exp(iQ_zT_z(i))} ] exp(i\overrightarrow{Q_{//}}\overrightarrow{R_{i}})
\sum _{n=0}^{N_c-1}\exp (iQ_{z}nL)
 \label{formfactorbundle}
\end{equation} 
where  $\overrightarrow{R_{i}}$ is the tube i position and $T_z(i)$ is a random number between 0 and L (see fig. 1).
Using the above procedure, one finds that the intensity
scattered by a powder of peapod \textit {bundles} is given by

\begin{displaymath}
I_{p}(Q)= \frac{(2\pi)^2 f_{s}^{2}}{QL^2} [ ( 2\pi r_{h}L\sigma
_{c}J_{0}(Qr_{h})+4\pi r_{C_{60}}^{2}\sigma _{c}\frac{sin(Qr_{C_{60}})}{
Qr_{C_{60}}})^{2}\sum_{i,j}J_{0}\left( QR_{ij}\right)\end{displaymath}
\begin{equation} 
+2N_{T}Int(QL/2\pi)(4\pi
r_{C_{60}}^{2}\sigma _{c}\frac{sin(Qr_{C_{60}})}{Qr_{C_{60}}})^{2}] \label{app1bundle}
\end{equation}
The lattice term is ($\sum_{i,j}J_{0}\left( QR_{ij}\right)$) where $R_{ij}$ is the distance between tubes $i$ and $j$
($R_{ij}=|\overrightarrow{R_{i}}-\overrightarrow{R_{j}}|$);
$N_{T}$\ is the number of tubes per bundle.  If one considers a distribution of tube
diameters inside bundles, or distribution of bundle sizes, one can
extrapolate the average procedures presented in ref. \cite{rols}.

\section{Appendix 2}

This appendix details some intensity calculations in the case of a powder of nanotube bundles, for 
incomplete filling of the nanotubes by 'long' (quasi-infinite) chains of fullerenes. The chains within each tube
are assumed to be sufficiently long to allow one to describe their scattered intensity with Dirac distributions. This assumption implies that the formula derived 
below cannot be used for too small filling rates.
\newline
Giving the site (n) of a molecule in tube (i), we define the filling factor of the site (i,n) by a function f(i,n) 
which is 1 if the site is occupied and zero otherwise.  To calculate  the intensity $I_p(Q)$ in the case 
of partial filling, one multiplies the C$_{60}$  term in eq. \ref{formfactorbundle} in appendix 1 by 
this  filling factor. \newline
One follows the orientational average procedure given in appendix 1 
but with $I(\vec{Q})$ replaced by its average over occupancies f(i,n). 
One finds 
\[I_{p}(Q)= \frac{(2\pi)^2f_{s}^{2}}{QL^2}\{\sum_{i,j}[ (2\pi r_{h}L\sigma
_{c}J_{0}(Qr_{h}))^{2}+<p(i)p(j)> (4\pi r_{C_{60}}^{2}\sigma _{c}\frac{sin(Qr_{C_{60}})}{
Qr_{C_{60}}})^{2}\]
\[
+2 <p(i)> (2\pi r_{h}L\sigma
_{c}J_{0}(Qr_{h}))(4\pi r_{C_{60}}^{2}\sigma _{c}\frac{sin(Qr_{C_{60}})}{
Qr_{C_{60}}}) ] J_{0}\left( QR_{ij}\right)\]
\begin{equation}+2\sum_{i}Int(QL/2\pi)<p(i)p(i)>(4\pi r_{C_{60}}^{2}\sigma _{c}\frac{
sin(Qr_{C_{60}})}{Qr_{C_{60}}})^{2}\}  \label{Ipipj}
\end{equation}

During the course of the calculation terms containing $<p(i)p(j)>$ arise, where 
p(i) is the filling factor of tube i :
$p(i)=\frac{1}{N_{c}(i)} \sum_{n=1}^{N_{c}(i)}f(i,n)$ where ${N_{c}(i)}$ is the number of C$_{60}$ sites 
in tube i.
 \newline
The filling rate of the sample, called $<p>$ or $p$, is the double average of $f(i,n)$:

\begin{equation}
<p>=\frac{1}{N_{T}N{c}}\sum_{i=1}^{N_{T}}\sum_{n=1}^{N_{c}}f(i,n)
\end{equation}
where the number of C$_{60}$ sites has been taken to be the same for all tubes $N_{c}(i)=N_{c}$.

Three cases are considered.\newline
(i){\it Homogeneous partial filling of the tubes with long chains of C$_{60}$, 
all tubes having the same mean filling rate p, independent of i }: 
$p(i)=p$, independent of i.\

In that case, equation \ref{Ipipj} becomes :

\[
I_{p}(Q)= \frac{(2\pi)^2f_{s}^{2}}{QL^2}[ (2\pi r_{h}L\sigma
_{c}J_{0}(Qr_{h})+ p 4\pi r_{C_{60}}^{2}\sigma _{c}\frac{sin(Qr_{C_{60}})}{
Qr_{C_{60}}})^{2}\sum_{i,j}J_{0}\left( QR_{ij}\right)\]
\begin{equation} +2N_{T}Int(QL/2\pi)(p 4\pi
r_{C_{60}}^{2}\sigma _{c}\frac{sin(Qr_{C_{60}})}{Qr_{C_{60}}})^{2}]
\label{app2hom}
\end{equation}

The C$_{60}$\ form factor -$4\pi r_{C_{60}}^{2}\sigma _{c}\frac{
sin(Qr_{C_{60}})}{Qr_{C_{60}}}$- is multiplied by the mean filling rate p.
There is no other modification of equation \ref{app1bundle} because one
assumes that the C$_{60\text{ }}$molecules agglomerate within nanotubes to
form long chains.

(ii){\it Partial filling of the tubes with long chains of C$_{60}$ molecules, with filling
rates different from one tube to another within each bundle}. In that case, one has to consider that
\newline $<p(i)p(j)>=<p^2> $ if $i=j$ \newline $<p(i)p(j)>=<p>^2$ if $i\ne j$ 
(no correlation from one tube to another). So that $<p^2>-<p>^2\ne 0$.\newline

For instance, one can consider within each bundle average proportions p of fully filled tubes and (1-p)
of empty tubes (which can be due to the fact that p\% of the tubes are opened and (1-p)\% are closed), which
gives: $<p(i)>$=p and $<p(i)p(i)>$=p, then $<p(i)^2>-<p(i)>^2=p(1-p)\not=0$.
\newline
The different fillings of nanotubes induce
additional disorder in direct space, which corresponds to diffuse scattering
in reciprocal space.\ The additional scattering is the most intense at small
Q values as in the case of chemical disorder \cite{guinier}.\ This approach
is detailed in ref.\cite{NTAFI} for partial filling of zeolite channels with
nanotubes. Eq. \ref{app1bundle} becomes 
\[I_{p}(Q)= \frac{(2\pi)^2f_{s}^{2}}{QL^2}[ (2\pi r_{h}L\sigma
_{c}J_{0}(Qr_{h})+<p>4\pi r_{C_{60}}^{2}\sigma _{c}\frac{sin(Qr_{C_{60}})}{
Qr_{C_{60}}})^{2}\sum_{i,j}J_{0}\left( QR_{ij}\right)\]
\begin{equation}
+N_{T}(<p^{2}>-<p>^{2})(4\pi r_{C_{60}}^{2}\sigma _{c}\frac{sin(Qr_{C_{60}})
}{Qr_{C_{60}}})^{2}+2N_{T}Int(QL/2\pi)<p^{2}>(4\pi r_{C_{60}}^{2}\sigma _{c}\frac{
sin(Qr_{C_{60}})}{Qr_{C_{60}}})^{2}]  \label{app2inhom}
\end{equation}

(iii) {\it Nanotubes in the same bundle are all filled or all empty, p\% of the
bundles corresponding to filled tubes and (1-p)\% to empty ones.} 
The scattered intensity is the sum of the intensities for p fully filled
bundles and for (1-p) empty bundles. Equation \ref{app1bundle} becomes :

\[I_{p}(Q)= p\frac{(2\pi)^2f_{s}^{2}}{QL^2}[ (2\pi r_{h}L\sigma
_{c}J_{0}(Qr_{h})+4\pi r_{C_{60}}^{2}\sigma _{c}\frac{sin(Qr_{C_{60}})}{
Qr_{C_{60}}})^{2}\sum_{i,j}J_{0}( QR_{ij})
\] 
\begin{equation}
+2N_{T}Int(QL/2\pi)(4\pi
r_{C_{60}}^{2}\sigma _{c}\frac{sin(Qr_{C_{60}})}{Qr_{C_{60}}})^{2}]
+(1-p)\frac{(2\pi)^2f_{s}^{2}}{QL^2}[ (2\pi r_{h}L\sigma
_{c}J_{0}(Qr_{h}))^{2}\sum_{i,j}J_{0}( QR_{ij})]
\label{app2kataura}
\end{equation}

\section{Appendix 3}

This appendix details calculations in the case of polymerized C$_{60}$
molecules inside the tubes. The chains within the tubes are assumed to be formed of n-polymers of C$_{60}$ 
molecules.
The distance between bonded C$_{60}$ molecules within a n-polymer is L$_{\text{b}}$\ , which is smaller than the distance L between
C$_{60}$ neighbors belonging to different n-polymers. For instance, for n=2, one considers a chain of dimers and for n=3 a chain of trimers.

The form factor of a n-polymer of C$_{60}$ molecules writes 
\[
F_{n-polymer}=4\pi r_{C_{60}}^{2}\sigma _{c}
{\displaystyle {sin(Qr_{C_{60}}) \over Qr_{C_{60}}}}
\left( 
\mathop{\displaystyle \sum }_{k=0}^{n-1}e^{iQ_{z}(k-\frac{(n-1)}{2})L_{b}}\right)\]
\[ =4\pi
r_{C_{60}}^{2}\sigma _{c}
{\displaystyle {sin(Qr_{C_{60}}) \over Qr_{C_{60}}}}
\left( e^{-iQ_{z}\frac{(n-1)}{2}L_{b}}\frac{1-e^{iQ_{z}nL_{b}}}{
1-e^{iQ_{z}L_{b}}}\right) =4\pi r_{C_{60}}^{2}\sigma _{c}
{\displaystyle {sin(Qr_{C_{60}}) \over Qr_{C_{60}}}}
\frac{sin(Q_{z}\frac{nL_{b}}{2})}{sin(Q_{z}\frac{L_{b}}{2})} 
\]

where (k-$\frac{(n-1)}{2})L_{b}$ is the position of the C$_{60}$\ molecule indexed by k within the n-polymer.

By replacing in eq. \ref{I1tube} the monomer form factor by that of the n-polymer and the period
L along the monomer chain by the one along the n-polymer chain, which
is (L+(n-1)L$_{b}$), one obtains:

$
I(\vec{Q})\propto f_{s}^{2}[(2\pi r_{h}(L+(n-1)L_{b})\sigma
_{c}J_{0}(Qr_{h})
\frac{\sin (Q_{z}(L+(n-1)L_{b})/2)}{ Q_{z}(L+(n-1)L_{b})/2}
+4\pi r_{C_{60}}^{2}\sigma _{c}\frac{sin(Qr_{C_{60}})}{Qr_{C_{60}}}\frac{
sin(Q_{z}\frac{nL_{b}}{2})}{sin(Q_{z}\frac{L_{b}}{2})})^{2}\delta
(Q_{z})$
\[
+(4\pi r_{C_{60}}^{2}\sigma _{c}\frac{sin(Qr_{C_{60}})}{Qr_{C_{60}}}
\frac{sin(Q_{z}\frac{nL_{b}}{2})}{sin(Q_{z}\frac{L_{b}}{2})})^{2}\sum_{k\neq
0}\delta (Q_{z}-2\pi k/(L+(n-1)L_{b}))] 
\]

Powder average thus gives :

\[
I_{p}(Q)\propto f_{s}^{2}[ \frac{1}{Q}(2\pi r_{h}(L+(n-1)L_{b})\sigma
_{c}J_{0}(Qr_{h})
+4\pi r_{C_{60}}^{2}\sigma _{c}n\frac{sin(Qr_{C_{60}})}{
Qr_{C_{60}}})^{2}\]
\[+\frac{2}{Q}(1-\delta _{N,0})
\mathop{\displaystyle \sum }%
_{k=1}^{M}(4\pi r_{C_{60}}^{2}\sigma _{c}\frac{sin(Qr_{C_{60}})}{Qr_{C_{60}}}
\frac{sin(\pi knL_{b}/(L+(n-1)L_{b}))}{sin(\pi k L_{b}/(L+(n-1)L_{b}))}
)^{2})] 
\]

\strut where M is the integer part of ($\frac{{\bf Q(L+(n-1)L}_{b}}{{\bf 2}
\pi }$); the term $(1-\delta _{M,0})$\ was introduced to avoid the case M=0.

\begin{figure}%[tbp]
\caption{Schematic representation of the system of coordinates and variables
used in the calculations.}\
%\end{figure}

%\begin{figure}%[tbp]
\caption{Calculations of the intensity diffracted by several powders. a)
nanotube of radius r=6.8 \AA \ and of length 380 \AA. b) linear chain of 40 C$
_{60}$ molecules. c) peapod (plain line), sum of the intensities from a) and
b) (dotted line), and crossed term (dashed line).}\
%\end{figure}

%\begin{figure}%[tbp]
\caption{Upper part of the figure : calculated powder diffraction patterns of
peapods for different tube radii : 5.42 \AA\ (8,8), 6.8 \AA\ (10,10), 8.1 \AA\ (12,12)
, with an inter-C$_{60}$ distance L equal to 9.5 \AA. Lower part of
the figure : calculated diffraction patterns for different inter-C$_{60}$
distances L and for a r=6.8 \AA\ tube.
}\
%\end{figure}

%\begin{figure}%[tbp]
\caption{Calculated diffracted intensities of a powder of isolated
nanotubes ($r_{h}$=6.8 \AA). Up: for random filling by the C$_{60}$
molecules (tube length is 380 \AA); down : for incomplete
filling by long ('infinite') C$_{60}$
chains. Filling rates indicated in the figure are the same for all tubes in each case considered.}\
%\end{figure}

%\begin{figure}%[tbp]
\caption{Component of the calculated diffracted intensities relative 
to periodicity effects (only the second term of eq. \ref{polym} 
is drawn) of a powder of isolated
nanotubes ($r_{h}$=6.8 \AA ), for C$_{60}$ monomers (upper line), dimers
(middle line) and trimers (lower line) chains inside the tubes.}

%\end{figure}

%\begin{figure}%[tbp]
\caption {Comparison between the calculated intensities scattered by a powder of
bundles of 12 empty nanotubes (upper curve of each part) 
and bundles of 12 peapods (lower curve of each part) for (10,10) nanotubes (upper part) of radius r=6.8 \AA, (8,8) nanotubes (middle part) of radius r=5.42 \AA\ and (12,12) nanotubes (lower part) of radius r=8.1 \AA. The tube length was fixed at 380 \AA\ in all the calculations. The arrows in the upper part
point toward assymetric peaks characteristic of the 1D periodicity of  C$_{60}$ chains.
}\
\end{figure}

\newpage
\begin{figure}%[tbp]
\caption{ {\it Left top} : Diffraction patterns of  85\% filled bundles of 12 peapods. {\it Left bottom} : Diffraction patterns of 50\% filled bundles of 12 peapods. All tubes have a radius of 6.8 \AA. {\it Right part} : Schematic representations of the 4 different filling modes used in calculations : a) random positions for the C$_{60}$ molecules within each tube, 
same filling for each tube, b) long C$_{60}$ chains, same filling for each tube , c) long C$_{60}$ chains, inhomogeneous 
filling of the different tubes within a bundle, d) mix of full and empty bundles.}\
 
%\end{figure}

%\begin{figure}%[tbp]
\caption{Effect on diffraction pattern of the filling of the external
channels with C$_{60}$s linear chains for different sizes of bundles. All tubes have a radius of 6.8 \AA\ and are 380 \AA\ long. a) 4
tubes. b) 6 tubes compared with a 6 tubes naked bundle (dotted line). c) 12
tubes.
}\
%\end{figure}

%\begin{figure}%[tbp]
\caption{Filling rate determination. a) First criterium giving the amplitude 
ratio I(.663\AA $^{-1}$)/I(.706\AA $^{-1}$) as function of the filling rate p 
(see the inset), calculated for randomly filled (empty circles) and homogeneously filled (full circles) bundles of peapods. b) Second criterium: amplitude ratio of the peaks around 
1.3 \AA $^{-1}$ and 1.1 \AA $^{-1}$ as function of p (see the inset), calculated for randomly filled (empty circles), inhomogeneously filled (empty triangles) and homogeneously filled (full circles) bundes of peapods. 
The dashed and dotted horizontal lines represent the experimental finding 
for the two investigated samples A and B.}\
%\end{figure}

%\begin{figure}%[tbp]
\caption{Experimental diffraction patterns 
of peapod samples. a) Peapod sample A (upper line), corresponding MER pristine SWNT
sample (lower line). The dotted line is the baseline used for the fits. b)
Peapods sample B (upper line) and a calculation of
its diffraction profile (lower line).}\
\end{figure}

\begin{thebibliography}{99}
\bibitem{noteL} The maximum of the asymmetric peak originating from the $C_{60}$ chains
 is situated at Q=2$\pi /L$
= 0.66 \AA $^{-1}$  for infinite chains. For finite chains, it becomes less abrupt and its 
maximum shifts to higher Q values 
(namely, 0.68 \AA $^{-1}$  for chains of 40 molecules in figure 3).
\end{thebibliography}
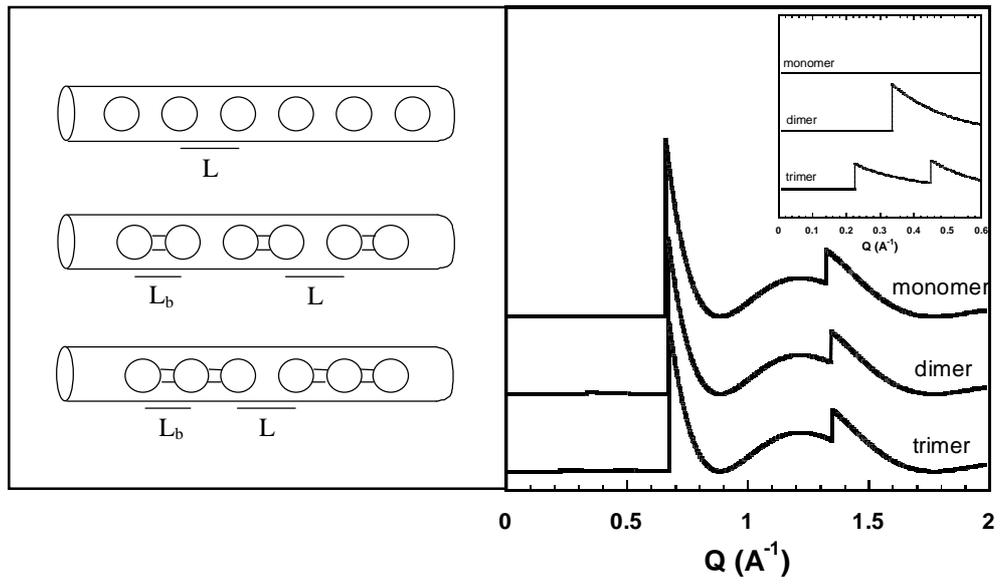
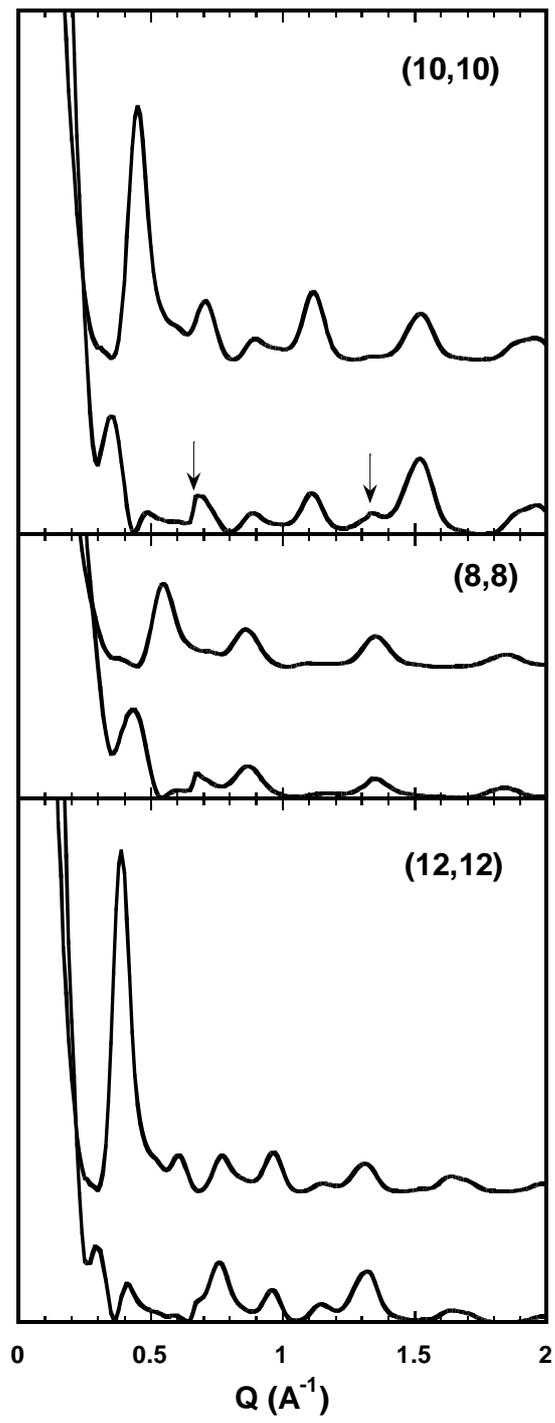
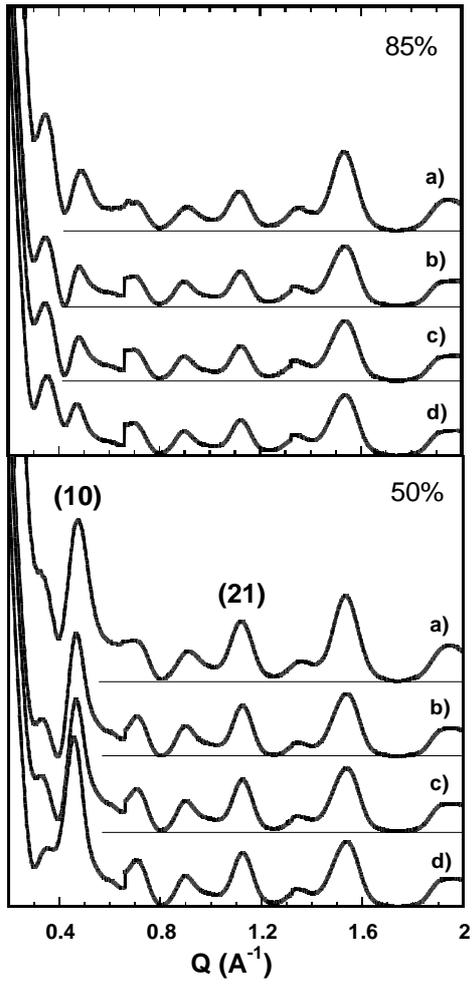
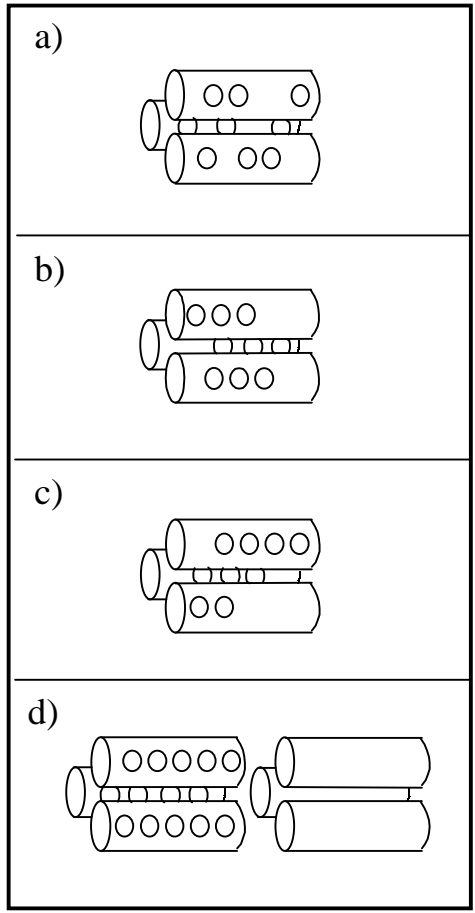
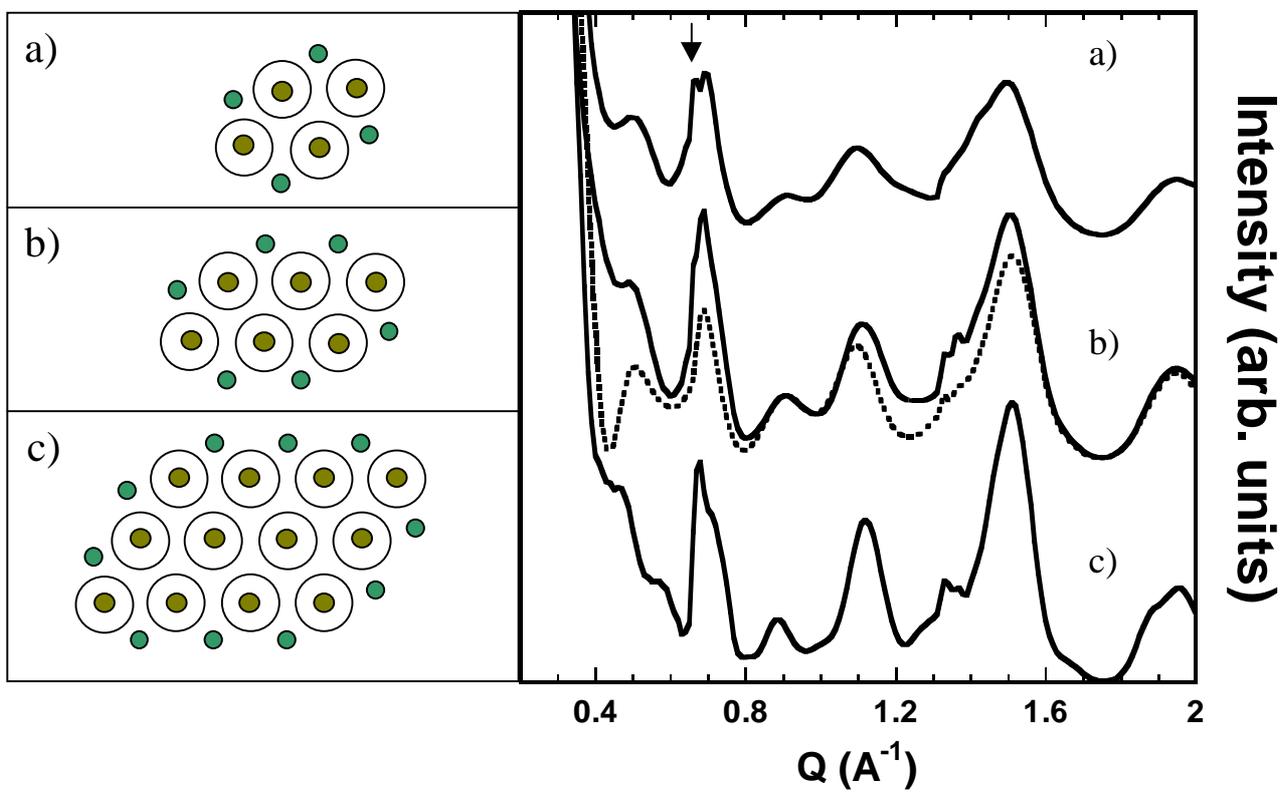
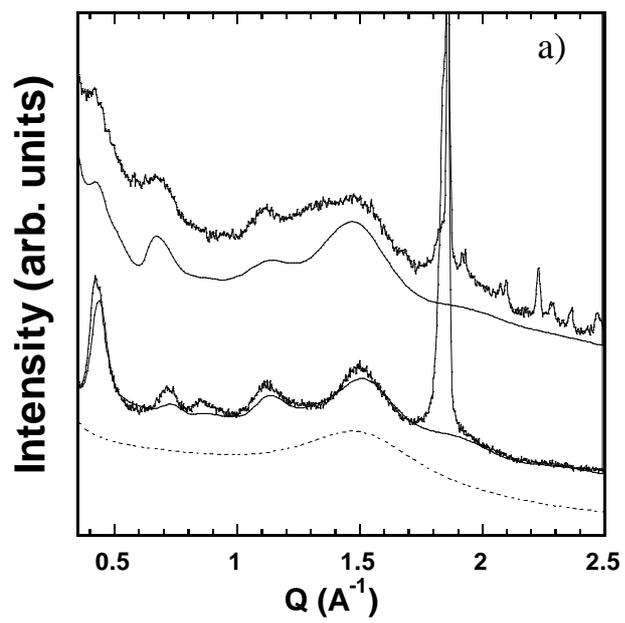
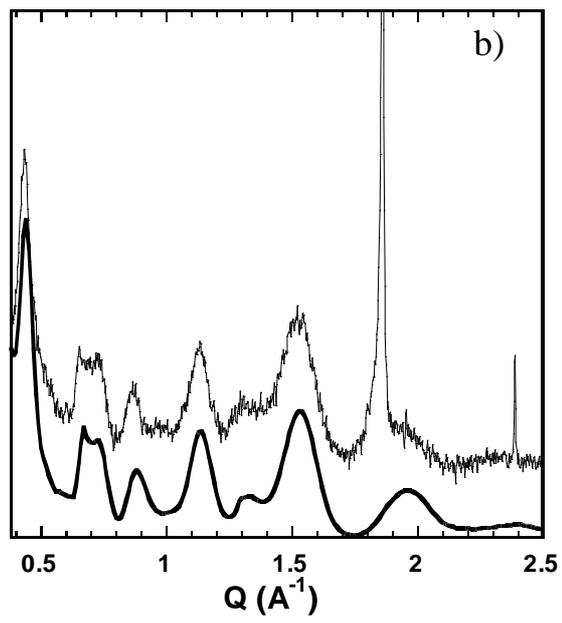
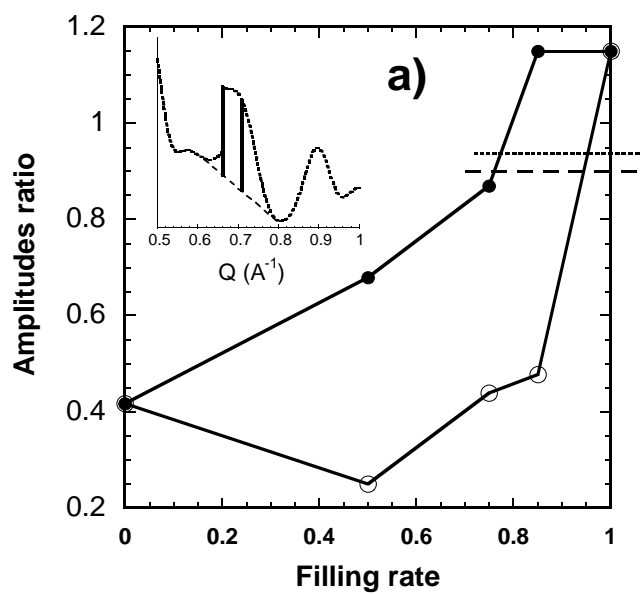
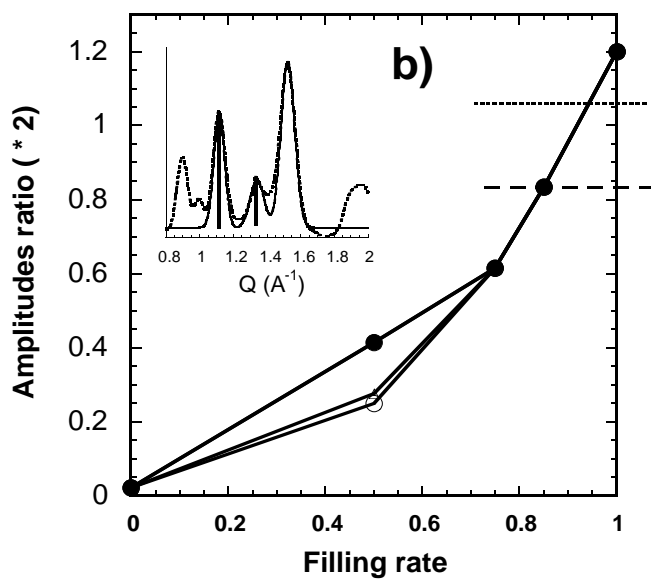
\end{document}